# A mid-infrared biaxial hyperbolic van der Waals crystal

Zebo Zheng,[1] Ningsheng Xu,[1] Stefano Luigi Oscurato,[2] Michele Tamagnone,[2] Fengsheng Sun,[1] Yinzhu Jiang,[1] Yanlin Ke,[1] Jianing Chen,[4,5] Wuchao Huang,[1] William L. Wilson,[3] Antonio Ambrosio,[3*] Shaozhi Deng,[1*] Huanjun Chen[1*]

**Affiliations:**

[1]State Key Laboratory of Optoelectronic Materials and Technologies, Guangdong Province Key Laboratory of Display Material and Technology, School of Electronics and Information Technology, Sun Yat-sen University, Guangzhou 510275, China.

[2]Harvard John A. Paulson School of Engineering and Applied Sciences, Harvard University, Cambridge, MA 02138, USA.

[3]Center for Nanoscale Systems, Harvard University, Cambridge, MA 02138, USA.

[4]Institute of Physics, Chinese Academy of Science and Collaborative Innovation Center of Quantum Matter, Beijing 100190, China.

[5]School of Physical Sciences, University of Chinese Academy of Science, Beijing 100049, China.

*Correspondence to: chenhj8@mail.sysu.edu.cn (H.C.); stsdsz@mail.sysu.edu.cn (S.D.); ambrosio@seas.harvard.edu (A.A.).

**Abstract:** Hyperbolic media have attracted much attention in the photonics community, thanks to their ability to confine light to arbitrarily small volumes and to their use for super-resolution applications. The 2D counterpart of these media can be achieved with hyperbolic metasurfaces, which support in-plane hyperbolic guided modes thanks to nanopatterns which, however, pose significant fabrication challenges and limit the achievable confinement. We show that thin flakes of the van der Waals material α-MoO$_3$ can support naturally in-plane hyperbolic polariton guided modes at mid-infrared frequencies without any patterning. This is possible because α-MoO$_3$ is a biaxial hyperbolic crystal, with three different Restrahlen bands, each for a different crystal axis. Our findings can pave the way towards new paradigm to manipulate and confine light in planar photonic devices.

**One Sentence Summary:** Van der Waals α-MoO$_3$ crystal has been demonstrated as a mid-infrared biaxial hyperbolic crystal, which supports in-plane phonon polaritons with concave wavefronts.

Hyperbolic media are characterized by a permittivity tensor having the component along one axis with opposite sign with respect to the other two axes. They have been extensively studied for their exotic optical properties, particularly because they can support electromagnetic fields with arbitrarily high momenta and hence achieve very high light confinement (*1−3*). This concept can be directly generalized to 2D media, where surface waves or guided waves are considered, and implemented using hyperbolic metasurfaces (HMSs) (*4−6*). These are flat photonic nanopatterned structures that support guided waves with in-plane hyperbolicity and can control the two-dimensional light propagation in unconventional way, giving rise to a variety of intriguing optical phenomena such as all-angle negative refraction (*7*), greatly enhanced photonic density of states (*4, 8*), and wavefronts with concave curvatures, to name a few (*9, 10*). The necessity of patterning, however, typically leads to strong optical losses, and limits the actual confinement that can be reached. Furthermore, the electromagnetic responses of the HMSs are governed by the permittivity tensors that are derived from the effective medium theory, which is only valid in the long-wavelength limit where the structural periodicity is much smaller than the incidence wavelength. Consequently, the hyperbolic dispersion is constrained to a very small region in the reciprocal space, leading to rather limited electromagnetic wavevectors. Such an issue can in principle be alleviated by reducing the structural periodicities of the metasurfaces down to sub-10 nm scale. However, this should happen without bringing in additional surface roughness or defects, which is a great challenge for nano-fabrication techniques (*11*). Thus, the quest for a natural medium that can be used to achieve in-plane hyperbolicity without nanopatterning is a very important open problem in nano-photonics.

Three-dimensional hyperbolic responses can be found in specific types of naturally existing homogeneous materials with strong dielectric anisotropy (*11−14*). Differently from an artificial hyperbolic material, the structural periodicities in a natural hyperbolic material are those of the crystal lattice, hence at an atomic scale. This results into a hyperbolic dispersion extending into a much larger region in the reciprocal space (*12, 15*). In particular, the van der Waals (vdW) crystals, which can be exfoliated layer-by-layer down to a single atomic planar layer, have been shown excellent hyperbolic properties from the UV to the THz (*11, 13−15, 16, 17*). However, to the best of our knowledge, all of the vdW materials reported so far are uniaxial crystals with negative (positive) and isotropic in-plane permittivity while positive (negative) out-of-plane permittivity. When such materials are exfoliated in 2D flakes, because the optical axis coincides with the exfoliation direction, they cannot achieve an in-plane hyperbolic response, unless patterned (similarly to HMSs) (*18*).

In a recent work we have demonstrated highly confined hyperbolic phonon polaritons (PhPs) in a new type of vdW semiconducting crystal, the alpha-phase molybdenum trioxide (α-$MoO_3$), grown by thermal physical deposition method (*19*). Now, we show that the vdW α-$MoO_3$ is actually a type of natural biaxial hyperbolic crystals with pristine in-plane hyperbolic dispersion in the mid-infrared regime. We use high-resolution optical scanning probe nano-imaging techniques to investigate phonon polariton modes launched, guided, and manipulated within the α-$MoO_3$ thin flakes in the different hyperbolicity bands of the crystal. In particular, the concave wavefront of a polaritonic mode originated from the in-plane hyperbolicity of the flake is shown as an unmistakable signature of the preserved three-dimensional hyperbolicity of this 2D material.

Similar to other polar vdW crystal like hexagonal boron nitride (h-BN), the infrared spectrum of the α-$MoO_3$ is governed by the phonon absorptions. The lattice of the α-$MoO_3$ is composed of octahedron unit cells with nonequivalent Mo–O bonds along the three principal crystalline axes (Fig. 1A, inset) (*20, 21*). Such a crystalline structure gives rise to rich phonon modes that are infrared-active along different crystalline directions (*20*). In each Reststrahlen band between the longitude (LO) and transverse optical (TO) phonon frequencies, a high reflectivity and negative real part of the permittivity, Re($\varepsilon$), are expected. In the α-MoO$_3$ there are three Reststrahlen bands in the mid-infrared regime between 545 cm$^{-1}$ and 1010 cm$^{-1}$. Specifically, the first one (545 cm$^{-1}$ ~ 851 cm$^{-1}$) originates from the in-plane phonon mode along the [001] crystalline direction (*y*-axis) of the α-MoO$_3$. The second and third bands are located in the regions 820 cm$^{-1}$ ~ 972 cm$^{-1}$ and 958 cm$^{-1}$ ~ 1010 cm$^{-1}$, originated from phonon modes along the [100] (*x*-axis) and [010] (*z*-axis) directions, respectively (*19, 20*). The crystal exfoliation direction to reduce α-MoO$_3$ in thin flakes is *z*. In these three bands the α-MoO$_3$ exhibits different permittivities along the three principal axes ($\varepsilon_x \neq \varepsilon_y \neq \varepsilon_z$, see supplementary materials, note S1), as expected for a biaxial crystal. Furthermore, in the full range 545 cm$^{-1}$ and 1010 cm$^{-1}$ there is always at least one component with negative Re($\varepsilon$) (Fig. 1A), so that the crystal is hyperbolic in the whole frequencies range, though along different directions for each of the Restrahlen bands. Using the permittivity tensor, the isofrequency surfaces of the α-MoO$_3$ can be calculated by applying the general Fresnel equation

$$\sum_{j=x,y,z} \frac{\varepsilon_j k_j^2}{n^2 - \varepsilon_j} = 0$$ (see supplementary materials, note S1) (*22*). To simplify the description only

the Re($\varepsilon$) are used in the calculations. Considering a α-MoO$_3$ thin flake with surface perpendicular to the [010] direction (*z*-axis), as shown in Fig. 1B, the isofrequency surface for the phonon polaritons in the three Reststrahlen bands are asymmetric hyperboloids. Specifically, in Band 1 where the $\varepsilon_y < 0$ and $\varepsilon_x \neq \varepsilon_z > 0$, the polaritonic modes exhibit simultaneously out-of-plane and in-plane hyperbolic dispersion. Similar situation applies for negative $\varepsilon_x$ in Band 2. In these two regions, strongly confined polaritonic modes exist in the α-MoO$_3$ flake, with in-plane directional propagation depending on the specific hyperbolicity band. In Band 3 with negative $\varepsilon_z$ and unequal positive $\varepsilon_x$ and $\varepsilon_y$, the out-of-plane dispersion is instead hyperbolic, whereas the in-plane one is elliptical (Fig. 1B), with consequent anisotropic propagation of the polaritonic mode in the *x*–*y* plane. These complex isofrequency surface behaviors are distinctly different from those of the already reported uniaxial hyperbolic media (e.g. the h-BN and hyperbolic metamaterials) with negative (positive) and isotropic in-plane permittivity and positive (negative) out-of-plane permittivity. In these conventional media, nanostructuring is required for the in-plane hyperbolic responses to occur (*18, 23*).

The in-plane propagation behavior of phonon polaritons in our material can be predicted by inspecting the isofrequency curves intersected by the *x*–*y* plane. Figure 1C shows the respective isofrequency curves at typical frequencies of the three Reststrahlen bands. For frequencies at 623 cm$^{-1}$ (Band 1) and 925 cm$^{-1}$ (Band 2), two-sheet dispersion curves can be observed. In contrast, the dispersion for the frequency at 1000 cm$^{-1}$ (Band 3) is elliptical (see also fig. S1). The in-plane hyperbolic responses can be revealed by calculating the polaritonic wave launched onto the surface of a 160-nm-thick α-MoO$_3$ flake by an electric dipole source (Fig. 1D, see details in supplementary materials, note S2, fig. S2). Polaritonic waves with evident hyperbolic shapes are seen in the electric field amplitude distributions at both 623 cm$^{-1}$ and 925 cm$^{-1}$ (Fig. 1, E and F). In addition, the real parts of the $E_z$ exhibit concave wavefronts with opening directions along the *y*- (623 cm$^{-1}$) and *x*-axis (925 cm$^{-1}$), respectively (Fig. 1, H and I). On the other hand, radial propagation of elliptical shape is observed at 1000 cm$^{-1}$ (Fig. 1, G and J). These dispersive behaviours can be more clearly visualized by plotting the Fourier transform (FT) of the Re($E_z$) images (Fig. 1, K–M), which reproduce the trajectories shown in Fig. 1C. From the FT distributions, one can see that the in-plane hyperbolic dispersions can result into very large wave vectors, i.e. large field confinements ($k_{PhPs}/k_0$) that enables focusing and manipulating of the mid-infrared electromagnetic waves at deep subwavelength scale. To be more specific, the polaritonic mode will propagate preferentially along a cone, with axes coincident with the directions normal to the isofrequency curves shown in Fig. 1. This means that the polaritonic wave propagation direction at 623 cm$^{-1}$ (925 cm$^{-1}$) is at $\theta = 34°$ ($\theta = 58°$) with respect to the y-axis (Fig. 1, E and F). These angles agree well with the theoretical values of 36° and 55°, derived from the equation

$$\theta(\omega) = \frac{\pi}{2} - \arctan\left(\sqrt{\varepsilon_y(\omega)}/i\sqrt{\varepsilon_x(\omega)}\right) \ (12).$$

According to the theoretical description above, the propagation angle $\theta$ is very sensitive to the frequency $\omega$, increasing (decreasing) as the frequency increases in Band 1 (Band 2) (fig. S3). We observe similar behavior in the calculated electric field distributions (fig. S2, B to G). As the frequency increases, the hyperboloid progressively opens and becomes flat at 820 cm$^{-1}$ (in Band 1) where the hyperbolic polariton changes into a polaritonic plane wave. By further increasing the frequency into the Band 2, the hyperboloid orientation changes from the y-axis to the x-axis, where a similar trend with the frequency can be observed: as the frequency increases, the hyperboloids progressively opens (fig. S2, E to G). These behaviors unambiguously highlight the tailoring of the hyperbolic polaritons in the α-MoO$_3$ flake by changing the excitation frequency.

To experimentally verify the α-MoO$_3$ as a natural material with in-plane hyperbolic responses, we used optical nano-imaging techniques to directly visualize the polaritonic modes supported on a α-MoO$_3$ flake. In a first experiment, a silver nanowire antenna (with length of 2.5 μm and width of 60 nm) was deposited on top, the (010) crystalline plane (the x–y plane), of a 220-nm-thick α-MoO$_3$ flake (Fig. 2A, supplementary materials, note S3). Upon illumination with p-polarized light, the localized plasmons of the nanoantennas can concentrate the optical fields to its extremities, with high enough momentum to launch a polaritonic wave into the α-MoO$_3$ surface. The polaritonic field $E_p$ (specifically the direct contribution as described in Ref. *16* and *17*) interferes with the local response due to the material polarizability $E_{in}$ (i.e. the material contribution in Ref. *16* and *17*) and generate interference fringes on the sample surface. The interference patterns are mapped by collecting the scattered field from the metallic tip of a scattering-type Scanning Near-field Optical Microscope (s-SNOM) (Fig. 2B, supplementary materials, note S4). As the result of this interference, the near-field amplitude image directly visualizes the wavefronts of the polaritonic mode launched by the in-plane silver antenna.

The near-field optical image for incidence light at 944 cm$^{-1}$ (Band 2) is shown in Fig. 2C ($\varepsilon_x = -0.98+0.08i$, $\varepsilon_y = 1.12+0.018i$, and $\varepsilon_z = 10.49+0.57i$). Interference fringes with concave shape can be easily observed, originated from the tip of the nanoantenna. The anomalous wavefront exhibits opening direction along the [100] direction (*x*-axis). In addition, the FT of the s-SNOM image (Fig. 2D) clearly shows a dispersion of hyperbolic shape, which is in excellent agreement with the isofrequency curves calculated by the Fresnel equation without any adjustable parameters (Fig. 1).

When the illumination frequency is changed to 980 cm$^{-1}$ (Band 3) with positive real parts of the in-plane permittivities ($\varepsilon_x = 0.67+0.04i$, $\varepsilon_y = 1.56+0.014i$, and $\varepsilon_z = -0.23+0.064i$), the concave shape of the polariton wavefront changes into a convex one (Fig. 2E). This results from the fact that the polaritons in the Band 3 have no in-plane hyperbolic responses, as evident from elliptically shaped FT distribution (Fig. 2F). The experimental s-SNOM images can be further corroborated by the numerical simulations with the permittivities at these two frequencies (see details in supplementary materials, note S5). As shown in Fig. 2, G and I, the simulation near-field distributions show excellent agreements with the experimental s-SNOM images, either for the fringe separations and curvatures. The opening angles and separations of the fringes also changes as a function of the illumination frequency (see theoretical calculations in supplementary materials, fig. S4).

In order to complete the description of the polaritonic features observed in the s-SNOM images of Fig. 2, C and 2B, close to the flake edge fringes with twice the periodicity of those close to the antenna are observed. Such fringes result from polaritonic modes coupled into the flake by the SNOM tip (round trip components): the polaritons are launched by the tip, reflected by the edge and again sampled by the tip (*16*).

In another experiment, we obtained the in-plane polaritonic dispersion relations $\omega(q)$, with $q$ being the polaritons wavevector. In this case, s-SNOM imaging was performed on a carefully selected α-MoO$_3$ flake with two orthogonal natural edges, where one of them is along the [001] (*y*-axis) direction (edge I) and the other is along the [100] (*x*-axis) direction (edge II) as determined from the micro-Raman spectroscopy (Fig. 3A, supplementary materials, note S6, fig. S5). For excitation frequency at 910 cm$^{-1}$ and 926 cm$^{-1}$ (Band 2), only fringes parallel to the [001] direction can be observed at the edge I. This is consistent with the directional propagation of the polaritonic mode resulting from the in-plane hyperbolic dispersion in this band, where the opening of the isofrequency curve points towards the [100] direction (Fig. 1C). Such a topology imposes the polaritons to propagate perpendicularly to the edge I instead of the edge II (fig. S6, A to E). As a result, only fringes associated with the polariton waves launched by the s-SNOM tip and reflected from the edge I can be collected (upper panel in Fig. 3, D and F).

The opposite situation can be observed for frequencies in Band 1 using imaging with Photo-induced Force Microscopy (*16, 17*), as shown in Fig. 3, C, E and G, where fringes only perpendicular to the [001] direction can be observed. In Band 3 instead, where $\varepsilon_x \neq \varepsilon_y > 0 > \varepsilon_z$, the in-plane elliptical isofrequency curve allows polaritons propagating in all directions. This results in fringes parallel to both edge I and II, as shown in Fig. 3D (lower panel) for illumination at 986 cm$^{-1}$ and 992 cm$^{-1}$. More specifically, the lack of circular symmetry in the isofrequency curve result in a different periodicity for the fringes orthogonal to edge I and II at the same illuminating wavelength (Fig. 3F and fig. S7). This effect is also evident from the simulation of Fig. 3B (see also supplementary materials, note S7, fig. S7).

From the polaritonic fringe periodicity it is possible to calculate the polariton wavelengths ($\lambda_p$) corresponding to the illumination frequency ($\omega$) (fig. S6) and the polaritonic wavevector ($q = 2\pi/\lambda_p$) (*13, 15*). As shown in Fig. 3H, the dispersion relation of the phonon polaritons in the α-MoO$_3$ exhibit obvious in-plane anisotropic behaviors. The Type I/II hyperbolic responses with positive/negative in-plane real part of the permittivities are shown to exist on the α-MoO$_3$ flake. Specifically, for illumination frequency between 780 cm$^{-1}$ and 800 cm$^{-1}$, Type II PhPs propagating only along the [001] direction can be observed, where the $\lambda_p$ ($q$) decreases (increases) as the $\omega$ becomes large (supplementary materials, fig. S9, movie S1). Similar trend applies to the frequency range of 820 cm$^{-1}$ ~ 950 cm$^{-1}$, where the PhPs propagate toward the [100] direction instead (supplementary materials, movie S2). However, if one changes the $\omega$ to the range of 820 cm$^{-1}$ ~ 950 cm$^{-1}$, Type I PhPs can be observed along both of the [001] and [100] directions (supplementary materials, movie S1 and S2), which exhibit reduced $q$ as the $\omega$ is increased. These experimental results agree well with the theoretical 2D false color plot of the complex reflectivity (see supplementary materials, note S8, fig. S8). Additionally, the electromagnetic confinement (defined by $q/k_0$) of the PhPs can be calculated according to the dispersion relation, which can be as high as ~ 87 at 953 cm$^{-1}$ and in very good agreement with the theory.

The unique hyperbolicity of α-MoO$_3$ flakes can be combined with artificial confinement of the polaritonic modes imposed by structuring the flake into specific geometric shapes (*24, 25*). As a simple demonstration, we fabricate an α-MoO$_3$ disc of 2-μm-diameter using focus ion beam and characterize its optical near-field properties (see supplementary materials, note S3) with the s-SNOM. At an excitation frequency within the Band 3 (994 cm$^{-1}$), the reflections from the disk edges of the polaritonic modes launched by the SNOM tip produces interference patterns of elliptical shape (Fig. 4A). In fact, although the flake shape is cylindrically symmetric, the ellipticity of the isofrequency curve in this Band dominates far from the edges (center of the disk). The measured polaritonic anisotropy (1.47), defined as $\lambda_{px}/\lambda_{py}$, is comparable to that reported (1.25) in a recent study but with a rather complex artificial heterostructure (*26*). For the same disk, when the illumination frequency is in the Band 2 (900 cm$^{-1}$), the interference pattern is characterized by fringes deformed along the *x*-axis (Fig. 4B), which is associated with the in-plane hyperbolic isofrequency curves opening towards the *x*-axis (Fig. 1C).

One of the most promising applications for hyperbolic media is sub-diffraction optical focusing by taking advantage of their highly directional polariton waves (*2, 12, 15*). However, in artificial HMSs the degree of focusing is usually limited by the relatively small maximum wavevector they can provide. Such a limitation can be overcome with the natural hyperbolic materials due to their atomic scale unit cells (*27*). Moreover, in comparison with other types of natural hyperbolic materials, the α-MoO$_3$ is anticipated to nourish the optical focusing due to its biaxial hyperbolicity. As a proof-of-concept demonstration, we calculate the guiding and focusing of mid-infrared light by a 300-nm-thick α-MoO$_3$ flake. To visualize the focusing spot, a 20-nm-thick metallic square slab (400 nm × 400 nm) was placed underneath the MoO$_3$ flake as a nanoantenna for launching the polaritonic mode (Fig. 4, C and D). Due to the in-plane anisotropic hyperbolic dispersion, the mid-infrared light can be guided and then focused, depending on the incidence frequency as well as polarization. For example, if the MoO$_3$ flake is illuminated with a *x*-polarized plane-wave at 914 cm$^{-1}$ where the opening direction of the hyperboloid is along the [100] direction (*x*-axis), the polaritonic waves are launched from the two slab edges that are along the *y*-axis (Fig. 4F). These waves subsequently propagate at a fixed angle with respect to the y-axis, giving rise to narrow strips orientating along the [001] axis (*y*-axis). The widths of these strips are ~ 80 nm (Fig. 4E), which is about 1/136 of the incidence wavelenght. In contrast, if one excites the sample at 776 cm$^{-1}$ using *y*-polarized light, the incidence light is focused into strips that are along the *x*-axis (Fig. 4, G and H). Such a sub-diffraction focusing behaviour is resulted from the biaxial hyperbolic responses of the α-MoO$_3$ and can be used in polarization optics in the mid-infrared as well as for efficient light trapping in device applications.

The homogenous biaxial hyperbolic vdW α-MoO$_3$ crystal revealed in our study offers the prospect of planar photonics without the need for complex nanopatterning, which is unavoidable in the 2D metasurfaces counterparts. The attainable wavevectors of the guided electromagnetic waves within the α-MoO$_3$ are therefore in-principle limited only by the atomic crystalline periodicity, giving rise to ultrastrong electromagnetic confinements. The confinements and manipulations of the electromagnetic fields at the nanoscale could be further enriched by introducing of sophisticated nanostructuring processes (Fig. 4). Additionally, the α-MoO$_3$ is also a type of semiconductor, which enables applications of the hyperbolic media in active optoelectronic devices in terms of their excellent electrical transportation characteristics, physical characteristics sensitive to external doping, and high optical-to-electrical conversion efficacy. The biaxial hyperbolic characteristics revealed in our current study can in-principle be generalized to other types of layered materials with different Restrahlen bands associated with the three optical axes, which can facilitate versatile applications by choosing the materials with appropriate hyperbolic bands. Our results therefore can pave the way towards new paradigm to manipulate and confine light in planar photonic devices.

**ACKNOWLEDGMENTS**

**Funding:** The authors thank Mr. Ruihui He and Professor Weiguang Xie from Jinan University for providing parts of the α-MoO$_3$ samples. The authors acknowledge support from the National Natural Science Foundation of China (grant nos. 51290271, 11474364, 11874407), the National Key Basic Research Program of China (grant no. 2013CB933601), the National Key Research and Development Program of China (grant nos. 2016YFA0203500), the Guangdong Natural Science Funds for Distinguished Young Scholars (grant no. 2014A030306017), Pearl River S&T Nova Program of Guangzhou (grant no. 201610010084), the Guangdong Special Support Program (grant no. 201428004), and the Strategic Priority Research Program of Chinese Academy of Science, (grant no. XDB30000000). This work was performed in part at the Center for Nanoscale Systems (CNS), a member of the National Nanotechnology Coordinated Infrastructure Network (NNCI), which is supported by the National Science Foundation under NSF award no. 1541959. M.T. acknowledges the support of the Swiss National Science Foundation (SNSF) grant no. 177836.

**Author contributions:** H.C., S.D., and N.X. conceived and initiated the study. Sample fabrication was performed by Y.J., Y.K., and W.H. Z.Z., A.A., S.L.O., M.T., and Y.J. performed the experiments and numerical simulations. A.A. and S.L.O. performed the PiFM measurements. F.S. contributed to the dispersion calculations. J.C. contributed to the experiments. H.C., S.D., N.X., A.A., and W.L.W. coordinated and supervised the work, discussed and interpreted the results. H.C. and Z.Z. co-wrote the manuscript with the input of all other co-authors.

**Competing interests:** Authors declare no competing interests.

**Data and materials availability:** All data is available in the main text or the supplementary materials.

**SUPPLEMENTARY MATERIALS**

Materials and Methods

Notes S1–S8

Figures S1–S9

Captions for movies S1 and S2

Tables S1

References (*28, 29*)

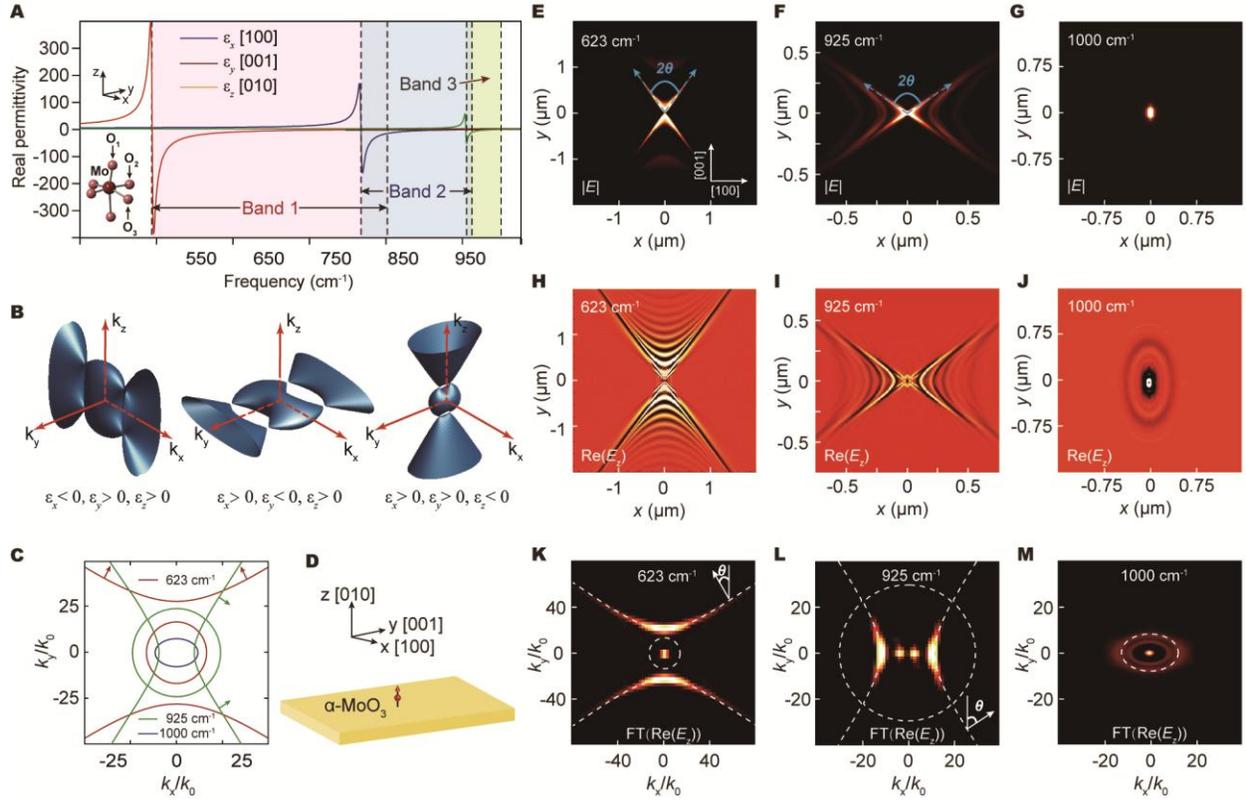

**Fig. 1. Mid-infrared biaxial hyperbolic electromagnetic responses of the vdW α-MoO₃ flake.**
(**A**) Real-part of the permittivities along the three principle axes. Different Reststrahlen bands are indicated in color. Inset: schematic showing the crystalline unit cell of the α-MoO₃. The three oxygen sites of asymmetric Mo–O bonds along different crystalline axes are indicated by $O_1$, $O_2$, and $O_3$, respectively. (**B**) 3D iso-frequency surfaces at different Reststrahlen bands, indicating a biaxial hyperbolic dispersion of the natural vdW α-MoO₃. (**C**) Iso-frequency curves on the $x$–$y$ plane at three typical frequencies. The red and green arrows indicate the propagation directions of the hyperbolic PhPs modes. (**D**) Schematic showing launching of the PhPs on the α-MoO₃ surface with a $z$-polarized electric dipole. (**E**–**G**) Calculated magnitudes of the electric field distributions, $|E|$, on the natural α-MoO₃ surface. (**H**–**J**) Calculated real parts of $z$-components of the electric field distributions, $Re(E_z)$, on the natural α-MoO₃ surface. Panels (H) and (I) demonstrate two hyperbolic PhPs with in-plane concave wavefronts, in contrast to the elliptical wavefront shown in panel (J). (**K–M**) Respective Fourier transforms (FT) of the panels (H)–(J). The white dashed lines indicate the corresponding isofrequency curves.

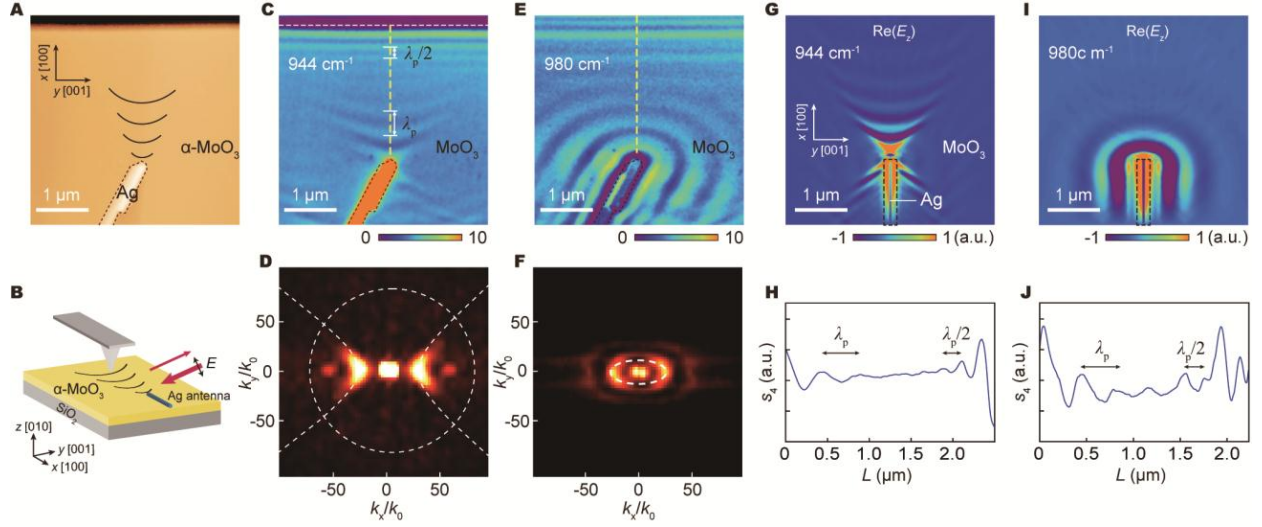

**Fig. 2. Real-space imaging of the hyperbolic PhPs on surface of a natural 220-nm-thick α-MoO₃ flake.** (**A**) Topography of the α-MoO₃ flake with silver nanoantenna placed onto its surface. (**B**) Schematic showing nano-imaging of the PhPs waves propagating on the α-MoO₃ surface. The black curves shown in panels (A) and (B) illustrate the wavefronts of the hyperbolic PhPs. (**C** and **E**) Real-space near-filed optical images recorded at 944 cm$^{-1}$ and 980 cm$^{-1}$, respectively. The edge of the flake and profile of the nanoantenna are respectively marked by the white and black dashed lines. (**D** and **F**) Absolute values of the FT images corresponding to panels (C) and (E), respectively. The white dashed lines indicate the theoretical isofrequency curves. (**G** and **I**) Numerical simulations of the Re($E_z$) distributions corresponding to the PhPs launched by the silver nanoantenna at 944 cm$^{-1}$ and 980 cm$^{-1}$, respectively. (**H** and **J**) Extracted profiles of the s-SNOM amplitudes along the yellow dashed lines indicated in panels (C) and (E), respectively.

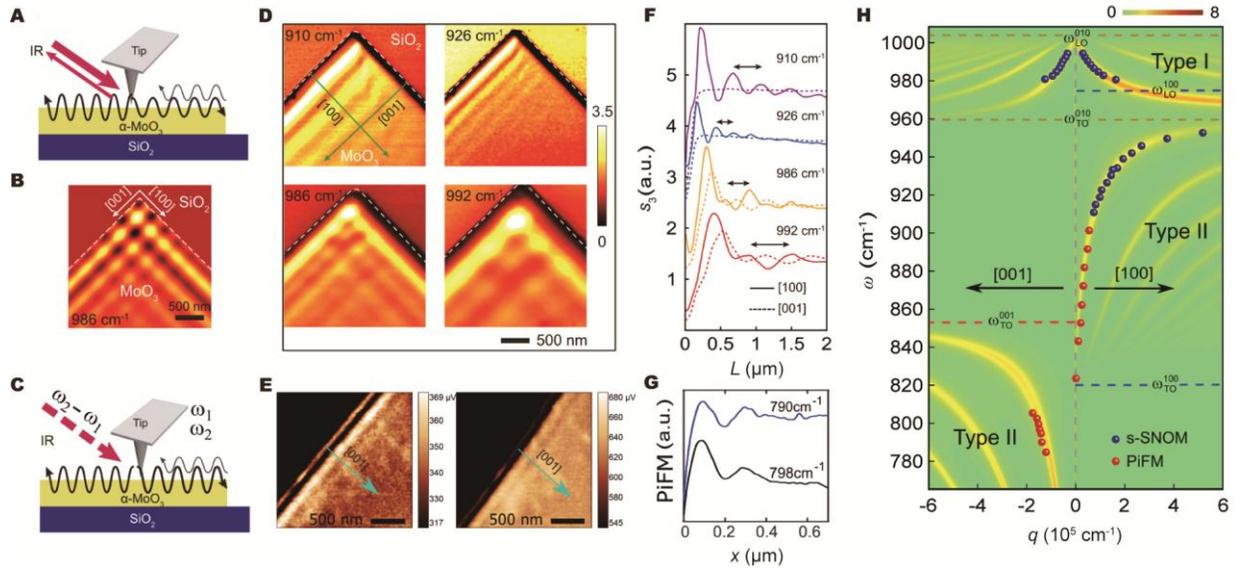

**Fig. 3. In-plane PhPs dispersion relations of a 160-nm-thick α-MoO₃ flake.** (**A**) Schematic showing real-space imaging of the PhPs reflected by the sample edges via s-SNOM. (**B**) Numerical simulation of the PhPs near-field distributions. (**C**) Schematic of PiFM imaging of the PhPs. Light is modulated at the difference frequency of the two first cantilever modes. (**D**) Near-field s-SNOM optical images obtained with varied illumination frequencies. (**E**) Near-field PiFM images obtained in the Band 1 at 790 and 798 cm$^{-1}$ (from left to right). (**F**) Near-field optical amplitude profiles along the [100] (solid) and [001] (dashed) directions extracted from panel (D). (**G**) PiFM profiles along the [001] direction extracted from panel (E). (**H**) Dispersion relation of the PhPs in the α-MoO₃ flake. Symbols indicate experimental data extracted from the s-SNOM (blue dots) and PiFM (red dots) images with different excitation frequencies. The false color image represents the calculated imaginary part of the complex reflectivity, Im$r_p(q, \omega)$, of the air/α-MoO₃/SiO₂ multilayer structure.

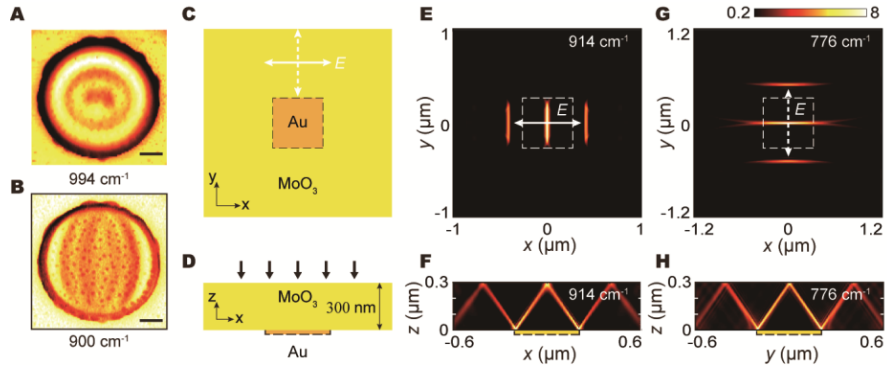

**Fig. 4. Sub-wavelength steering and focusing of the mid-infrared electromagnetic fields with the biaxial hyperbolic PhPs of the vdW α-MoO₃ flake.** (**A** and **B**) Near-field optical images of a 2 μm-diameter α-MoO₃ disc at different illumination frequencies. The thickness of the disc is 20 nm. (**C** and **D**) Schematic showing the architecture for the sub-wavelength electromagnetic focusing with the hyperbolic PhPs, where a gold plate is placed under the α-MoO₃ flake. The thicknesses of the gold plate and α-MoO₃ flake are 20 nm and 300 nm, respectively. (**E** and **G**) Calculated electric near-field distributions on the top surface of the α-MoO₃. (**F** and **H**) Calculated electric near-field distributions in the cross-section perpendicular to the top surface of the α-MoO₃ flake. The illumination frequencies are 914 cm$^{-1}$ (panels E and F) and 776 cm$^{-1}$ (panels G and H), respectively.

# Supplementary Materials for

## A mid-infrared biaxial hyperbolic van der Waals crystal

Zebo Zheng,[1] Ningsheng Xu,[1] Stefano Luigi Oscurato,[2] Michele Tamagnone,[2] Fengsheng Sun,[1] Yinzhu Jiang,[1] Yanlin Ke,[1] Jianing Chen,[4,5] Wuchao Huang,[1] William L. Wilson,[3] Antonio Ambrosio,[3*] Shaozhi Deng,[1*] Huanjun Chen[1*]

*Correspondence to: chenhj8@mail.sysu.edu.cn (H.C.); stsdsz@mail.sysu.edu.cn (S.D.); ambrosio@seas.harvard.edu (A.A.).

**Affiliations:**
[1]State Key Laboratory of Optoelectronic Materials and Technologies, Guangdong Province Key Laboratory of Display Material and Technology, School of Electronics and Information Technology, Sun Yat-sen University, Guangzhou 510275, China.
[2]Harvard John A. Paulson School of Engineering and Applied Sciences, Harvard University, Cambridge, MA 02138, USA.
[3]Center for Nanoscale Systems, Harvard University, Cambridge, MA 02138, USA.
[4]Institute of Physics, Chinese Academy of Science and Collaborative Innovation Center of Quantum Matter, Beijing 100190, China.
[5]School of Physical Sciences, University of Chinese Academy of Science, Beijing 100049, China.

*Correspondence to: chenhj8@mail.sysu.edu.cn (H.C.); stsdsz@mail.sysu.edu.cn (S.D.); ambrosio@seas.harvard.edu (A.A.).

**This PDF file includes:**

    Materials and Methods
    Supplementary Text
    Figs. S1 to S9
    Captions for movies S1 and S2
    Table S1

**Other Supplementary Materials for this manuscript include the following:**

    Movies S1 and S2

**Materials and Methods**

**Note. S1. Theoretical model for calculating the isofrequency surface of the PhPs**

*Anisotropic permittivities of the α-MoO₃:* In the mid-infrared region, the optical response of the α-MoO₃ is dominated by the phonon absorption rather than electronic transition, the dielectric function can be described by the following Lorentz equation *(13, 15)*:

$$\varepsilon_j = \varepsilon_\infty^j \left(1 + \frac{\omega_{LO}^{j\,2} - \omega_{TO}^{j\,2}}{\omega_{TO}^{j\,2} - \omega^2 - i\omega\Gamma^j}\right), \quad j = x, y, z \tag{S1}$$

where $\varepsilon_j$ denotes the principal components of the permittivity tensor. The $\varepsilon_\infty^j$ is the high frequency dielectric constant, the $\omega_{LO}^j$ and $\omega_{TO}^j$ refer to the longitude (LO) and transverse (TO) optical phonon frequencies, respectively. The parameter $\Gamma^j$ is the broadening factor of the Lorentzian lineshape. The *x*, *y*, and *z* denote the three principal axes of the crystal, which correspond to the crystalline directions [100], [001], and [010] of the α-MoO₃, respectively. To calculate the relative permittivities shown in Fig. 1A, the phonon frequencies are directly adopted from the literature values *(14, 15)*. The detailed parameters used in our calculation are given in Table S1:

*Fresnel equation of biaxial crystal:* we start from the Maxwell equations,

$$\nabla \cdot \vec{D} = 0 \tag{S2}$$

$$\nabla \cdot \vec{H} = 0 \tag{S3}$$

$$\nabla \times \vec{E} = -\mu_0 \frac{\partial \vec{H}}{\partial t} \tag{S4}$$

$$\nabla \times \vec{H} = \frac{\partial \vec{D}}{\partial t} \tag{S5}$$

where $\vec{D} = \vec{\varepsilon} \cdot \vec{E}$. Considering a monochromatic plane wave propagating in the crystal with a wavevector *k*, the corresponding *E*, *D*, and *H* can be stated as

$$\vec{E} = \vec{E}_0 \exp[i(\vec{k} \cdot r) - \omega t] \tag{S6}$$
$$\vec{D} = \vec{D}_0 \exp[i(\vec{k} \cdot r) - \omega t] \tag{S7}$$
$$\vec{H} = \vec{H}_0 \exp[i(\vec{k} \cdot r) - \omega t] \tag{S8}$$

By substituting the plane wave solution into the Maxwell equations we can obtain the following relations: $k \times E = \omega\mu_0 H_0$, $k \times H = -\omega D$. Note that $k = |k|k_0 = (\omega n/c)k_0$ (*n* denotes the refractive index of the medium, and $k_0$ is the unit vector of the wavevector), and the velocity of light $c = 1/\sqrt{\varepsilon_0\mu_0}$, therefore the electric displacement vector can be determined as

$$\vec{D} = -\frac{1}{\mu_0\omega^2} \vec{k} \times (\vec{k} \times \vec{E}) \tag{S9}$$

$$\vec{D} = -\varepsilon_0 n^2 \vec{k}_0 \times (\vec{k}_0 \times \vec{E}) = \varepsilon_0 n^2 [\vec{E} - \vec{k}_0(\vec{k}_0 \cdot \vec{E})] \tag{S10}$$

For a biaxial crystal, the three permittivity components are different from each other,

$$\vec{D}_j = \frac{\varepsilon_0 n^2 \vec{k}_{0j}(\vec{k}_0 \cdot \vec{E})}{\frac{n^2}{\varepsilon_j} - 1} = \frac{\varepsilon_0 \vec{k}_{0j}(\vec{k}_0 \cdot \vec{E})}{\frac{1}{\varepsilon_j} - \frac{1}{n^2}} \tag{S11}$$

The Maxwell equtions give $\vec{k} \cdot \vec{D} = 0$, thereby one can obtain,

$$\vec{D}_x \vec{k}_{0x} + \vec{D}_y \vec{k}_{0y} + \vec{D}_z \vec{k}_{0z} = 0 \tag{S12}$$

By substituting the Equation S11 into S12, one can get

$$\frac{\varepsilon_0 \vec{k}_{0x}^2 (\vec{k}_0 \cdot \vec{E})}{\frac{1}{\varepsilon_x} - \frac{1}{n^2}} + \frac{\varepsilon_0 \vec{k}_{0y}^2 (\vec{k}_0 \cdot \vec{E})}{\frac{1}{\varepsilon_y} - \frac{1}{n^2}} + \frac{\varepsilon_0 \vec{k}_{0z}^2 (\vec{k}_0 \cdot \vec{E})}{\frac{1}{\varepsilon_z} - \frac{1}{n^2}} = 0 \tag{S13}$$

Finally, we can obtain the Fresnel equation which describes the isofrequency surface of guided electromagnetic waves within the biaxial crystal as

$$\sum_{j=x,y,z} \frac{\varepsilon_j k_j^2}{n^2 - \varepsilon_j} = 0 \tag{S14}$$

## Note. S2. Simulation of the dipole-launched PhPs waves on the α-MoO₃ surface

To calculate the electric filed distributions (Fig. 1E to G) and wavefronts (Fig. 1H to J) of the dipole-launched PhPs, we performed the finite-difference time-domain (FDTD) simulations using a commercial software, the FDTD Solution (Lumerical, Canada). As shown in Fig. S2, an electric dipole (frequency range: 500 cm$^{-1}$ ~ 1250 cm$^{-1}$) is fixed above the α-MoO₃ flake supported on the SiO₂ substrate. The separation between the dipole source and α-MoO₃ surface is $h = 20$ nm. The thicknesses of the α-MoO₃ and SiO₂ layers are 160 nm and 300 nm, respectively. The anisotropic permittivities of the α-MoO₃ (Equation S1) are imported into the FDTD solutions as user-defined sample data. The magnitudes of the electric field distributions |E| and real parts of the z-components Re($E_z$) at typical frequencies are extracted at the top surface of the α-MoO₃ flake.

## Note. S3. Sample preparation

*Growth of the α-MoO₃ vdW crystal flakes:* The α-MoO₃ vdW flakes were synthesized by thermal physical deposition method *(28)* using a tube furnace. The MoO₃ powder of purity 99.9% (0.1 g) as source was placed at the center of a quartz tube. The cleaned SiO₂ substrate was placed at the low temperature zone 13.0 ~ 14.5 cm away from the source. The tube was heated up to 780 ℃ in 70 min and then kept at that temperature for 120 min. During the thermal treatment the MoO₃ powder was sublimated and recrystallized onto the low-temperature regions of the SiO₂ substrate. Subsequently the quartz tube was cooled down naturally to room temperature, whereby the α-MoO₃ vdW flakes with various thicknesses (tens to hundreds of nanometers) can be found on the SiO₂ substrate.

*Peraparation of the nanoantenna/α-MoO₃ heterostructre: Preparation of the nanoantenna onto the α-MoO₃ surface:* The high aspect ratio silver nanorods (with diameter of 50 nm ~ 60 nm and length of 2 μm ~ 3 μm) were synthesized using the overgrowth method, with the gold nanobipyramids as seeds *(29)*. The longitudinal plasmon resonance frequency of the silver nanoantenna can be precisely tuned by controlling their aspect ratios. In our current study, the plasmon frequency of the silver nanoantenna was ~ 1000 cm$^{-1}$. To integrate the nanoantenna onto the α-MoO$_3$ surface, an aqueous solution of the silver nanorods was drop-casted onto the SiO$_2$/Si substrate with randomly distributed α-MoO$_3$ flakes. Various nanoantennas can be found onto the α-MoO$_3$ surfaces after the droplet was dried naturally under ambient conditions.

*Fabrication the α-MoO$_3$ disk:* To fabricate the α-MoO$_3$ disk, we utilized a 30-keV focused ion beam (FIB) etching system (AURIGA, ZEISS). The beam diameter was kept at ~ 10 nm and the ion current was 5 pA. A piece of 150-nm-thick α-MoO$_3$ flake was chosen. The dose of the Ga$^{3+}$ beam for the etching was set at 0.6 nC/μm$^2$, with a dwell time of 0.3 μs. After the FIB fabrication, the sample was annealed at 300 ℃ for 2h in ambient condition to eliminate the intercalated Ga$^{3+}$ inside the α-MoO$_3$, whereby the PhPs characteristics can be recovered.

## Note. S4. Near-field optical imaging

The near-field optical measurements were conducted using a scattering-type near-field optical microscope (s-SNOM, NeaSNOM, Neaspec GmbH), which was built on the basis of an atomic force microscope (AFM). The near-field optical distributions of the sample can be mapped simultaneously with its topography. To image the PhPs in real-space, a mid-infrared laser (QCL, Daylight solutions) with tunable frequencies from 900 cm$^{-1}$ to 1240 cm$^{-1}$ was focused onto the sample through a metal-coated AFM tip (Arrow-IrPt, Nanoworld). During the measurements, the AFM was operated in tapping mode, where the tip was vibrated vertically with a frequency of $f$ = 280 kHz. The back-scattered light from the tip was collected by a MCT detector (HgCdTe, Kolmar Technologies). The near-field signal was extracted by applying pseudoheterodyne interferometric method, the detected signal was demodulated at high harmonic n$f$ (n ≥ 3) of the tip vibration frequency.

## Note. S5. Simulation of the nanoantenna-launched PhPs on the α-MoO$_3$

The FDTD solution was employed to calculate the wavefronts of the PhPs waves launched by the silver nanoantenna. As shown in Fig. S4, a silver nanoantenna is placed onto the surface of the α-MoO$_3$ flake supported by the SiO$_2$ substrate. The longitudinal axis of the antenna is parallel to the [100] crystalline direction of the α-MoO$_3$. The diameter and length of the nanoantenna is 60 nm and 2.5 μm, respectively. The thicknesses of the α-MoO$_3$ and SiO$_2$ layers are 220 nm and 300 nm, respectively. Such a heterostructure is illuminated vertically by a broad-band plane-wave, with polarization parallel to the longitudinal axis of the nanoantenna. The real part of $z$ components of the electric field, Re($E_z$), are obtained from the top surface of the α-MoO$_3$.

## Note. S6. Determination of the crystalline direction of a typical α-MoO$_3$ flake

The orthorhombic α-MoO$_3$ is a layered oxide composed of distorted and edge-shared MoO$_6$ octahedra (space group of *Pbnm*) (Fig. S5A). The asymmetric oxygen species along the different crystalline directions give rise to the various Mo–O bonding types, resulting in a number of vibration modes that are either IR active or Raman active. As shown in Fig. S5C, the Raman peak locating at 158 cm$^{-1}$ depicts the vibration of the translation rigid chain along the [001] direction, and the peak at 820 cm$^{-1}$ is originated from the O–Mo–O stretching vibration along the [100] direction. The peak at 996 cm$^{-1}$ is due to the stretching vibration of the Mo–O bond along the [010] direction *(19–21)*.

To accurately identify the crystalline directions of a specific α-MoO$_3$ flake, we carried out the polarization-revolved Raman spectroscopy. In a specific measurement, the Raman spectra were consecutively taken upon varying the angles $β$ between the edge I (Fig. S5B) of the flake and the polarization direction of the incidence laser. Figure S5C shows the Raman spectra taken with varied excitation polarizations. It is clear that the intensity of the peak at 158 cm$^{-1}$, which corresponds to the lattice vibration along the [001] direction, is strongly dependent on the excitation polarization. The intensity maximum appears at $β = 0°$ where the polarization of the incidence laser is parallel to the longitudinal axis of the flake. As the $β$ is increased, the Raman intensity decreases, which finally disappears at $β = 90°$ (Fig. S5D). As a result, the longitudinal direction of the flake is [001] direction, while the transverse one is [100] direction.

## Note. S7. Simulation of the s-SNOM image using the phenomenological cavity model

The optical near-field distributions on the vdW α-MoO$_3$ flake can be calculated using a phenomenological cavity model, where the PhPs are launched by a point source (AFM tip) located on the α-MoO$_3$ surface. The PhPs will propagate radially until reflected by the boundaries. The reflected waves from the boundaries will return to the source point and interfere with the subsequently tip-launched wave. The total field below the point source can be described as *(19, 24)*,

$$\psi = \tilde{\psi}_{PhPs,0} + \sum_j \tilde{\psi}_{PhPs,j} \tag{S15}$$

where $\tilde{\psi}_{PhPs,0}$ is the tip-launched PhPs wave. The reflected waves from the boundary $j$ can be described as $\tilde{\psi}_{PhPs,j} = R_j \times \tilde{\psi}_{PhPs,0} \exp[-(2q)r_j(\gamma+i)]$. Parameter $R_j = R_0 \exp(i\Delta\varphi)$ denotes the complex reflection coefficient, which includes the reflectivity $R_0$ and phase shift $\Delta\varphi$ of the wave. The parameters $\gamma$ and $r_j$ describe the polariton damping rate and distance between the flake edge and AFM tip, respectively. The interference patterns are associated with $|\psi|$. On the basis of this model, one can calculate the local filed amplitude beneath the AFM tip at arbitrary position on the α-MoO$_3$ sheet.

In the s-SNOM measurement, the AFM tip not only excites the PhPs by focusing the incidence mid-infrared light onto the α-MoO$_3$ flake, but also acts as a near-field detector for recording the total field beneath the tip. In the calculations the amplitude of the tip-launched polariton $\tilde{\psi}_{PhPs,0}$ was set to 1. For calculating the image shown in the Fig. 3B, the damping rate $\gamma$ was fixed at 0.12 and the values of $q$ can be calculated through the following equation $q(\omega) + iq'(\omega) = \dfrac{2}{d \tan\theta}\Delta\varphi$ *(13, 15)*, where $\Delta\phi = \arctan(\dfrac{\varepsilon_0}{\varepsilon_\perp \tan\theta}) + \arctan(\dfrac{\varepsilon_{sub}}{\varepsilon_\perp \tan\theta})$. Note that the in-plane wavevector $q$ is strongly depend on the propagation direction. To calculate the $q$ along the [001] direction ([100] direction), the $\varepsilon_\perp$ is substituted by $\varepsilon_y$ ($\varepsilon_x$). The $\varepsilon_{sub}$ denotes the relative permittivity of the SiO$_2$ substrate. For the complex reflection coefficient, the $R_0$ and $\Delta\varphi$ were set as 1 and 1.5π, respectively.

## Note. S8. Calculations on the complex reflectivity of the multilayer structure α-MoO$_3$/SiO$_2$

The PhPs dispersion relation of the α-MoO₃ flake can be manifested by the complex reflectivity $r_p(\omega, q)$ of a multilayer structure consisted of air/α-MoO₃/SiO₂ (Fig. S8). The total reflectivity of this system is determined by Equation S16 *(13)*.

$$r_p = \frac{r_1 + r_3 e^{i2k_2^b d}}{1 + r_1 r_3 e^{i2k_2^b d}} \tag{S16}$$

$$r_1 = \frac{\varepsilon_\perp k_1^b - \varepsilon_1 k_2^b}{\varepsilon_\perp k_1^b + \varepsilon_1 k_2^b} \tag{S17}$$

$$r_3 = \frac{\varepsilon_3 k_2^b - \varepsilon_1 k_3^b}{\varepsilon_3 k_2^b + \varepsilon_\perp k_3^b} \tag{S18}$$

where the subscripts "1", "2", and "3" denote the air, α-MoO₃, and SiO₂, respectively. The parameter $d = 160$ nm is the thickness of the α-MoO₃ flake. The parameters $r_1$ and $r_2$ delegate the reflectivities at the air/α-MoO₃ and α-MoO₃/SiO₂ interfaces, respectively. The $\varepsilon_j$ ($j = 1, 3,$ and $\perp$) refers to the relative permittivities of the air, SiO₂, and the component perpendicular to the *z*-axis ([010] direction) of the α-MoO₃ flake. To calculate the dispersion relation of the PhPs propagating along the [001] direction ([100] direction), the $\varepsilon_\perp$ is substituted by $\varepsilon_y$ ($\varepsilon_x$), which are obtained from the Equation S1. The parameter $k_j^b$ is the momentum component of the propagating wave along the *b* axis. For $j = 1$ and 3, the momenta are given by

$k_j^b = \sqrt{\varepsilon_j \left(\frac{\omega}{c}\right)^2 - q^2}$ , while for the electromagnetic wave in the anisotropic α-MoO₃

$k_2^b = \sqrt{\varepsilon_\perp \left(\frac{\omega}{c}\right)^2 - \frac{\varepsilon_\perp}{\varepsilon_{//}} q^2}$ .

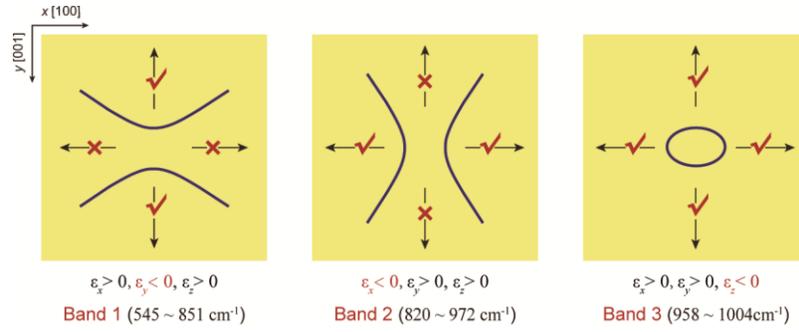

**Fig. S1. Schematic showing the isofrequency curves of the PhPs in the biaxial α-MoO₃ flake.** The opening directions of the hyperbolic isofrequency curves can be tuned by varying the illumination frequencies within the Band 1 (left) and Band 2 (middle), indicating the PhPs can propagate along the direction where the permittivity is negative. The right panel shows conventional in-plane anisotropic PhPs with elliptical wavefront, indicating the PhPs can propagate in all directions, but with different wavevectors.

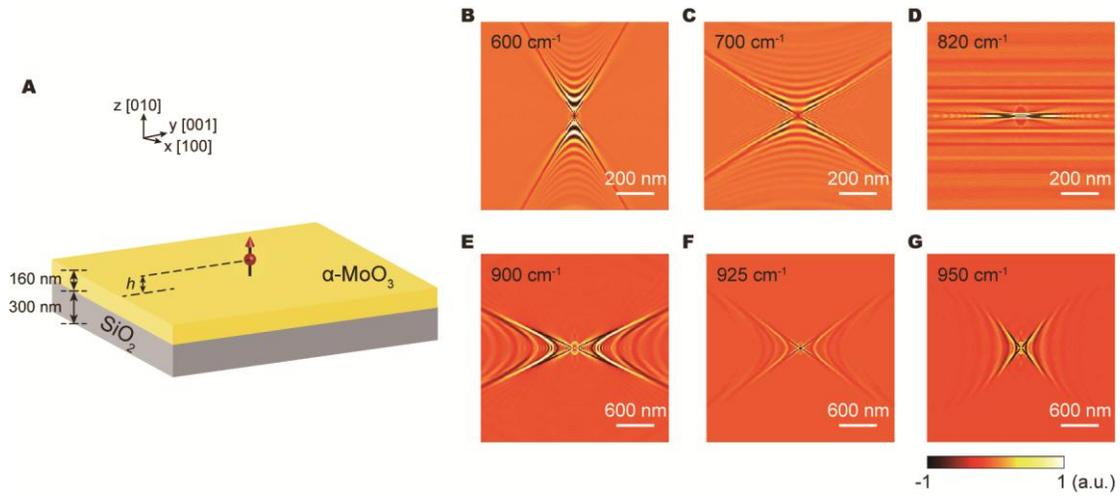

**Fig. S2. Simulations of the dipole-launched hyperbolic PhPs on the α-MoO₃ surface**. (**A**) Calculation scheme. The thicknesses of the α-MoO₃ flake and SiO₂ substrate are 160 nm and 300 nm, respectively. The $h$ = 20 nm denotes the separation between the electric dipole and α-MoO₃ surface. (**B**–**G**) In-plane hyperbolic PhPs with different wavefronts and propagation directions at various excitation frequencies.

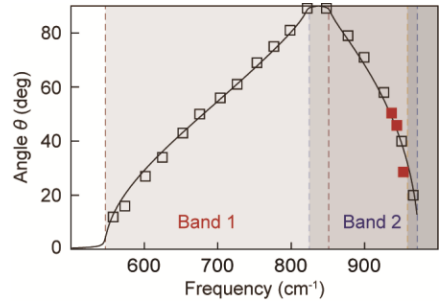

**Fig. S3. Propagation angle of the in-plane hyperbolic PhPs.** The solid line plots the theoretical calculation using equation $\theta(\omega) = \frac{\pi}{2} - \arctan\left(\sqrt{\varepsilon_y(\omega)}/i\sqrt{\varepsilon_x(\omega)}\right)$, the hollow squares are extracted data from the FDTD simulations, and the red solid squares are data obtained from the near-field optical images of shown in Fig. 2

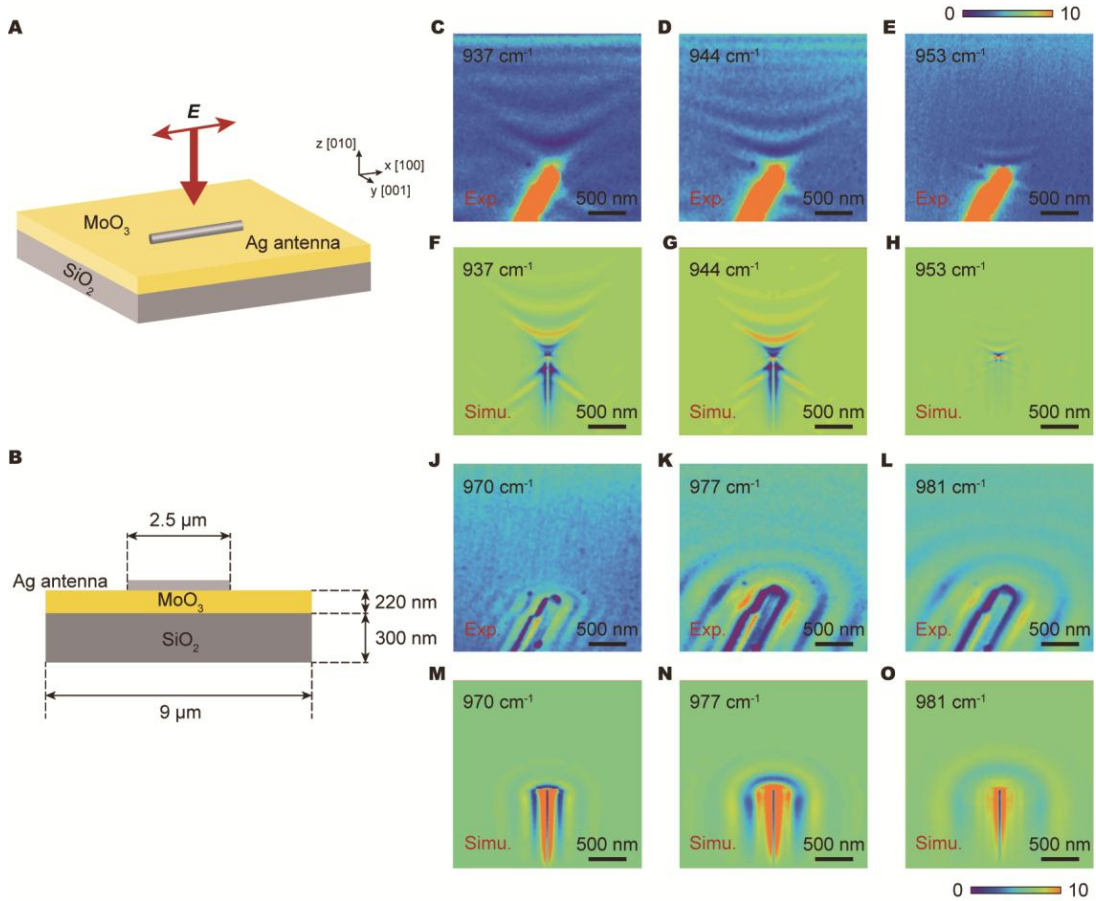

**Fig. S4. Silver nanoantenna-launched PhPs at various incidence frequencies.** (**A**) Schematic showing the simulation geometry. (**B**) Side-view of the nanoantenna/α-MoO₃/SiO₂ heterostructure. (**C–E**) Near-field optical images of the in-plane hyperbolic PhPs launched by the silver nanoantenna. The illumination frequencies locate in the Band 2. (**F–H**) Simulated electric field distributions of the nanoantenna-launched PhPs, which correspond to (C) – (E). (**J–L**) Near-field optical images of the PhPs excited with frequencies in the Band 3. (**M–O**) Simulated electric field distributions of the nanoantenna-launched PhPs, which correspond to (J) – (L).

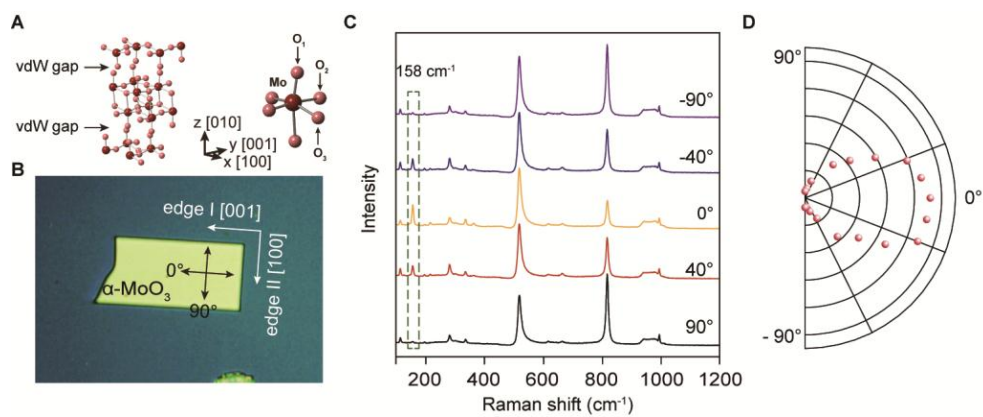

**Fig. S5. Determination of the crystalline directions of the α-MoO₃ by Raman spectroscopy.** (**A**) Crystalline structure of the α-MoO₃ and the corresponding octahedron unit cell. (**B**) Optical image of a typical α-MoO₃ flake with two orthogonal edges, where one of them is along the [001] direction (edge I) and the other is along the [100] direction (edge II). The black double-headed arrows indicate the polarization directions of the incidence laser, which is defined by the angle $\beta$ with respect to the [001] direction. (**C**) The Raman spectra of the α-MoO₃ obtained at different $\beta$. (**D**) Polar plots of the Raman intensities at 158 cm$^{-1}$.

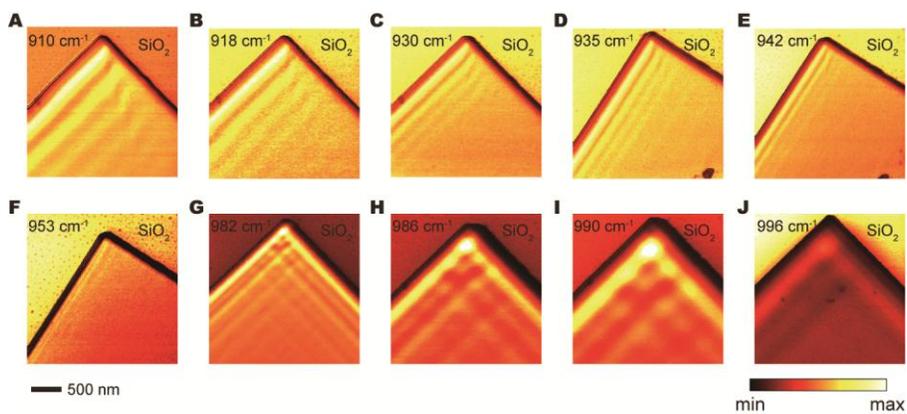

**Fig. S6. Near-field optical images showing in-plane anisotropic PhPs characteristics of the α-MoO₃.** (**A–F**) Near-field optical images obtained with different illumination frequencies in the Band 2, where the in-plane hyperbolic responses take place. (**G–J**) Near-field optical images obtained in the Band 3 without the in-plane hyperbolic responses.

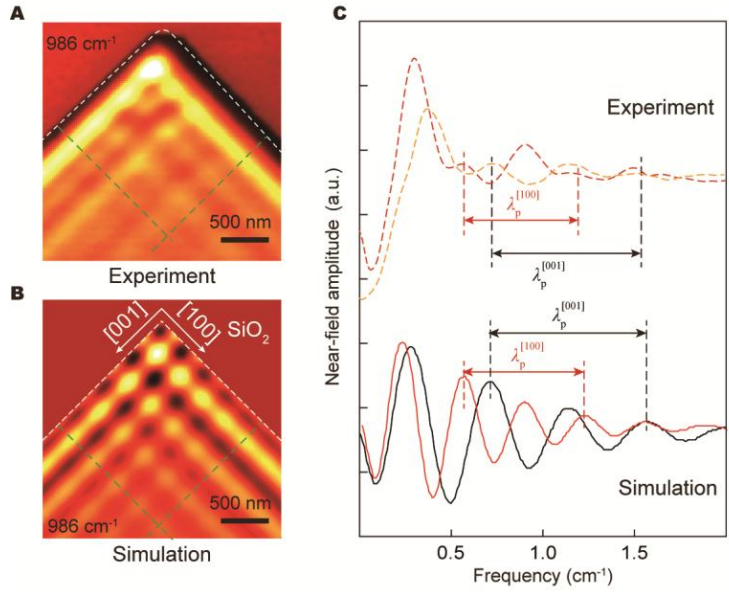

**Fig. S7. Comparison of the experiment and simulation near-field images of the PhPs distributions illuminated by 986 cm$^{-1}$ (Band 3).** (**A**) Near-field optical image obtained using the s-SNOM. (**B**) Simulated image using the phenomenological cavity model described in the note S7. (**C**) Comparison of the extracted profiles along the green dash lines marked in (A) and (B).

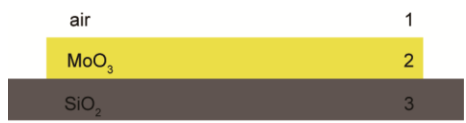

**Fig. S8. Scheme of the multi-layered structure consisted of air/α-MoO₃/SiO₂.** The characters "0", "1", and "2" denote the air, α-MoO$_3$, and SiO$_2$, respectively.

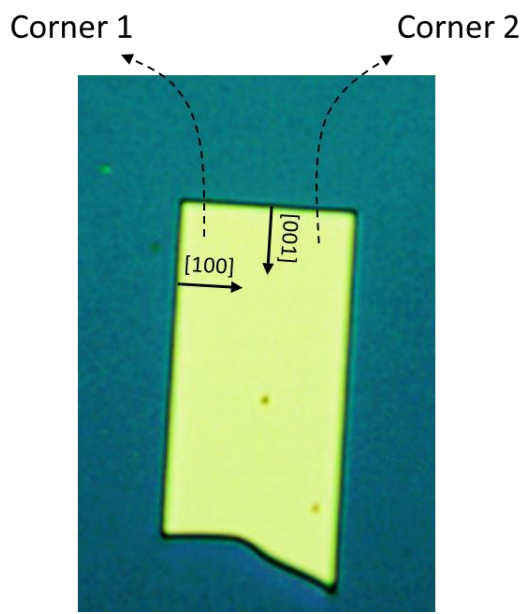

**Fig. S9. Optical image of the α-MoO$_3$ flake utilized for conducting the hyperspectral PiFM movies.**

| Direction | x [100] Ref (3, 4) | x [100] this work | y [001] Ref. (3, 4) | y [001] this work | z [010] Ref. (3, 4) | y [010] this work |
|---|---|---|---|---|---|---|
| $\varepsilon_\infty$ | \ | 4.0 | \ | 5.2 | \ | 2.4 |
| $\omega_{LO}$/cm$^{-1}$ | 974 | 972 | 851 | 851 | 1010 | 1004 |
| $\omega_{TO}$/cm$^{-1}$ | 818 | 820 | 545 | 545 | 962 | 958 |
| $\Gamma$/cm$^{-1}$ | \ | 4 | \ | 4 | \ | 2 |

**Table S1. Parameters used for calculating the permittivities (equation S1) in our study.**

**MOVIES S1 and S2**

**Movie S1 Hyperspectral PiFM movie of the edge perpendicular to the [100] direction, which is adjacent to the Corner 1 shown in Fig. S9.** The scan size is 3 μm ×3 μm. The spectral range is 776 cm$^{-1}$ ~ 1000 cm$^{-1}$.

**Movie S2 Hyperspectral PiFM movie of the edge perpendicular to the [001] direction, which is in proximity of the Corner 1 shown in Fig. S9.** The scan size is 3 μm ×3 μm. The spectral range is 776 cm$^{-1}$ ~ 1000 cm$^{-1}$.